\documentclass[12pt]{article}

\usepackage{scicite}

\usepackage{times}

\usepackage{amsmath}
\usepackage{amsfonts}
\usepackage{amssymb}
\usepackage{graphicx}
\usepackage{url}

\usepackage{booktabs}
\usepackage{subfig}

\usepackage{todonotes}

\newcommand{\cat}[1]{\textit{#1}}

\newenvironment{sciabstract}{
\begin{quote} \bf}
{\end{quote}}

\title{Estimating Traffic Disruption Patterns with Volunteered Geographic Information} 

\author
{ Chico Q. Camargo${}^{1}$, Jonathan Bright${}^{1}$, Graham McNeill${}^{1}$,\\Sridhar Raman${}^{2}$, Scott A. Hale${}^{1,3\ast}$\\
\\
\normalsize{${}^{1}$Oxford Internet Institute, University of Oxford, Oxford, United Kingdom}\\
\normalsize{${}^{2}$Oxford Brookes University, Oxford, United Kingdom}\\
\normalsize{${}^{3}$Alan Turing Institute, London, United Kingdom}\\
\\
\normalsize{$^\ast$To whom correspondence should be addressed; E-mail: scott.hale@oii.ox.ac.uk}
}

\date{}

\begin{document} 

\baselineskip=24pt

\maketitle 

\vspace{-0.5cm}

\begin{sciabstract}
Accurate understanding and forecasting of traffic is a key contemporary problem for policymakers. Road networks are increasingly congested, yet traffic data is often expensive to obtain, making informed policy-making harder. This paper explores the extent to which traffic disruption can be estimated from static features from the volunteered geographic information site OpenStreetMap (OSM). We use OSM features as predictors for linear regressions of counts of traffic disruptions and traffic volume at 6,500 points in the road network within 112 regions of Oxfordshire, UK. We show that more than half the variation in traffic volume and disruptions can be explained with static features alone, and use cross-validation and recursive feature elimination to evaluate the predictive power and importance of different land use categories. Finally, we show that using OSM's granular point of interest data allows for better predictions than the aggregate categories typically used in studies of transportation and land use. 
\end{sciabstract}

\section*{Introduction}

Understanding and forecasting traffic is an important task for urban policymakers. Road networks are by far the most heavily used part of transport infrastructure (for example, $64\%$ of all trips in the UK were made by car in 2016~\cite{transportbritain}); yet compared to other transportation modes (such as rail and air) basic data about traffic flow on roads is largely lacking. 
In the last decade, a variety of novel data sources have started to offer the possibility of filling this gap, such as data from GPS transponders on mobile phones (see ref.~\cite{vlahogianni2014short} for a review) or data from social media~\cite{mcneill2017estimating}, which are generating considerable academic interest. Here, we contribute to this growing literature on the use of new data sources to understand traffic by using volunteered geographic information from OpenStreetMap (OSM) to understand what types of land use are associated with traffic jams, as well as increased traffic volume. 

The connection between land use and transport is a classic subject in the literature~\cite{wegener2004land,Lenormand2015,Louail2015}, though land use categories are often classified at a highly aggregate level (e.g., defining areas as \cat{residential}, \cat{commercial}, or \cat{industrial}) and data have typically been expensive to put together~\cite{liu2012urban}. OSM is very promising in this regard in that its data is highly granular, offering a classification of different types of commercial activity, public amenities and other forms of land use, but also in the fact that all this data is freely and openly available. The completeness and accuracy of OSM coverage has been assessed in previous studies \cite{haklay2010good,girres2010quality,zielstra2010comparative,helbich2012comparative,mashhadi2015impact,arsanjani2015quality,senaratne2017review,bright2018geodemographic,bright2018openstreetmap}, yielding positive but cautious results, particularly about road networks. It has also been used to successfully identify the types of trips which human mobility models struggle to predict accurately \cite{camargo2019diagnosing}.

\section*{Results}

We test the extent to which OSM data can offer a good estimation of the volume of overall traffic and the number of traffic disruptions, defined as any deviation from normal smooth traffic on a road network, by making use of a series of linear regression models.
For the models of the traffic disruptions volume, observations are the geographic (latitude and longitude) points where traffic disruptions were observed in the network and the response variable is the number of traffic disruptions observed during the month of March 2017. 

\begin{figure*}[tb]
\centering
\includegraphics[width=\linewidth]{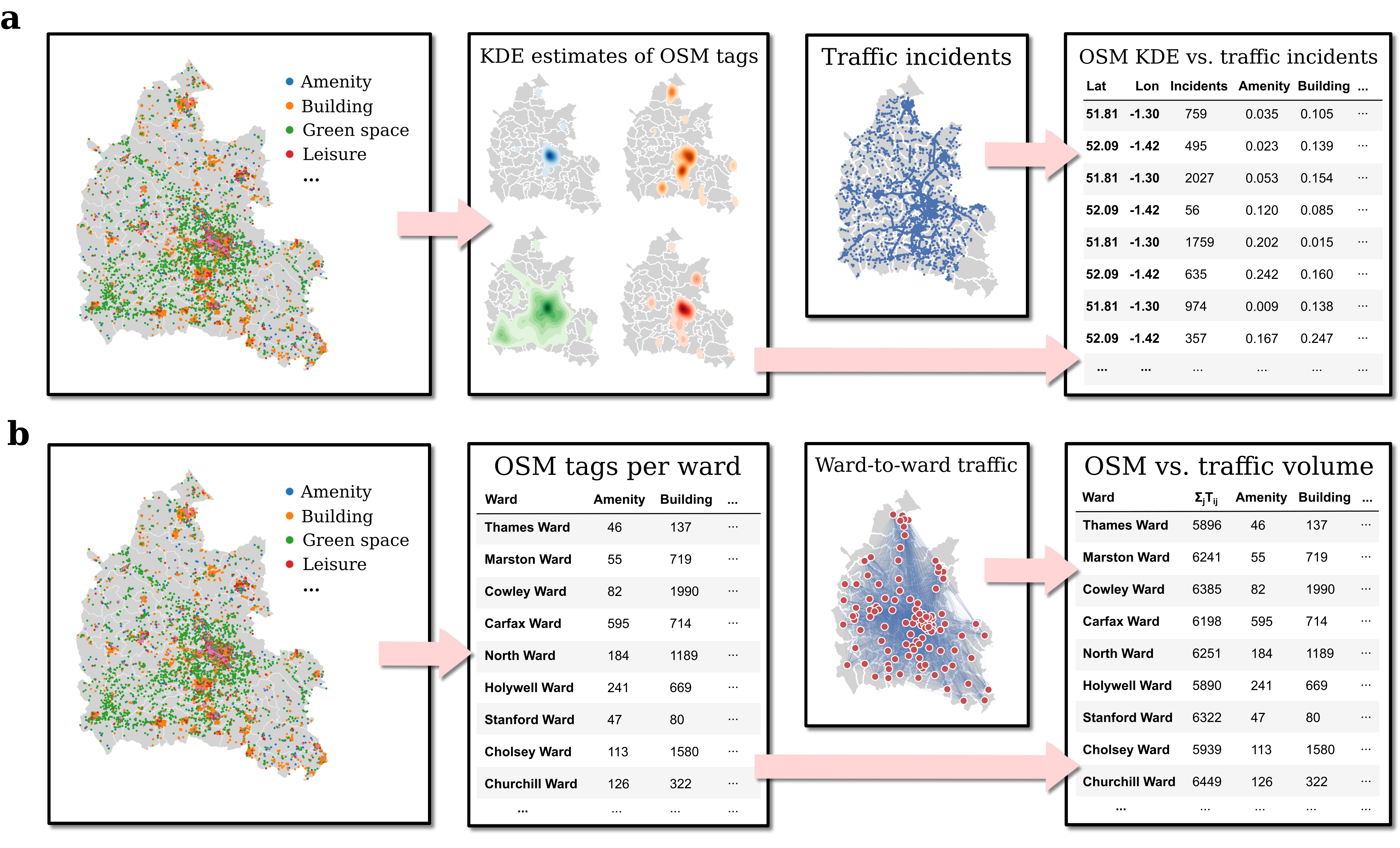}
\caption{\label{fig:drawing}
Schematic pipeline of the linear model for the two sets of linear models in this study.  As shown in the top panels \textbf{(a)}, we first we produce kernel density estimates (KDE) of every OpenStreetMap (OSM) category and meta-category, which we then compare with the number of traffic disruptions at a given latitude and longitude. The bottom panels \textbf{(b)} show we also aggregate the OSM data points into a total count per ward, which we then compare with the traffic volume going into every ward in Oxfordshire. 
}
\end{figure*}

The data analysis pipelines for the two sets of linear models in this study are described in Figure~\ref{fig:drawing}. As shown in the top panels \textbf{(a)}, we first produce kernel density estimates (KDE) of every OSM category and meta-category. We then estimate the number of traffic disruptions at a given latitude and longitude using the KDEs of either the OSM meta-categories or of the OSM categories at each point. To produce the KDEs, we made use of a Gaussian kernel searched over a range of bandwidth parameters before adopting a bandwidth of $0.001$, which captures the range of spatial variation of all OSM points of interest. The specific value of the bandwidth parameter did not qualitatively affect our results. These KDEs allow us to estimate the density of any type of OSM feature at all of the points where traffic disruptions were reported. 

As shown in the bottom panels \textbf{(b)}, we also perform a second set of linear regressions where we aggregate the OSM data points into a total count for every one of the 112 electoral wards in the county of Oxfordshire, UK. We then estimate the volume of traffic going into every ward using either counts of the OSM meta-categories or all OSM categories for each ward.

\begin{table}[tb]
    \centering
    \subfloat[Meta-categories only]{
        \begin{tabular}{lr}\toprule
            Variable & Estimate \\ \midrule
            Residential & -0.09** \\
            Industrial & -0.18** \\
            Recreational & -0.10* \\
            Institutional & 0.14* \\
            Green space & 0.26*** \\
            Commercial & 0.32*** \\ \midrule
            Observations & 6529 \\
            Adjusted $R^2$ & 0.11 \\ \bottomrule
        \end{tabular}
    }\hspace{5ex}
    \subfloat[Granular model]{
        \begin{tabular}{lr}\toprule
            Variable & Estimate \\ \midrule
            Residential & 0.61*** \\
            Farmland & 0.56*** \\
            Meadow & 0.18*** \\
            \ldots \\
            Cafe & -0.07* \\
            Apartments & -0.09** \\ \midrule
            Observations & 6529 \\
            Adjusted $R^2$ & 0.55 \\ \bottomrule
        \end{tabular}
    }
    \caption{Granular land-use categories from OpenStreetMap allow for more detailed understandings of traffic disruptions. Compared with the traditional land-use categories shown in (a) that produce an adjusted $R^2=0.11$, the granular classifications used in (b) increase the adjusted $R^2$ to $0.55$. Only a small subset of the 40 predictor variables are shown for (b). Respectively, *, ** and *** indicate $p<0.05$, $p<0.01$ and $p<0.001$. } 
    \label{tbl:linreg}
\end{table}

\subsection*{Estimating traffic disruptions}
The first linear model to estimate traffic disruptions only makes use of the meta-categories of OSM features (see Table~\ref{tbl:linreg}a). These meta-categories represent traditional classifications of land use types. The model only weakly fits the traffic disruptions data, resulting in an adjusted $R^2$ of $0.11$. 
Individual coefficients show that commercial areas are the ones most associated with high traffic, whilst industrial areas are the least so.
We also tested different versions of the model only estimating distributions on weekdays and weekends, as the nature of traffic disruptions on these days could be different, but the overall fit to the log-transformed data was similar.

The second model has more granular land-use data by making use of all OSM categories that were observed at least a hundred times in Oxfordshire, resulting in KDEs for 40 different types of point (from pubs, schools and restaurants to graveyards, postboxes and gardens). This model fits the log-transformed data considerably better than the meta categorization model as captured by the adjusted $R^2$, which is a goodness-of-fit metric that takes into account the different number of independent variables and is a common metric for model comparison in computational social science \cite{choi2012predicting,wu2015future,lin2019forecasting}. This granular model results in an adjusted $R^2 = 0.55$. The model coefficients of largest absolute value are represented in Table~\ref{tbl:linreg}b, and their corresponding p-values are indicated as well.

The second, granular model gives estimates of how things we might expect to explain local traffic jams vary with actual traffic disruptions. For example, one would expect places of worship and schools to both have a relatively high number of traffic disruptions, but the coefficients in this model indicate a large difference between the coefficient corresponding to the relationship between the number of points of interest tagged as schools and the log-transformed number of traffic disruptions and the corresponding coefficient for places of worship. The analysis, however, is only correlational: OSM points of interest tagged as farmland, parking and graveyards all have high positive coefficients. The high number of traffic disruptions around such points might be due to traffic network features such as narrow roads rather than the effects of these OSM features directly.

\subsection*{Estimating traffic volume}
We also test the effectiveness of OSM data in estimating the traffic volume in Oxfordshire. For this variable, rather than using KDEs to estimate the density of each OSM feature at a given road, we aggregate the number of points of interest tagged with each meta-category and category, producing two sets of independent variables for every ward: one corresponding to the total number of points tagged with each one of the 6 OSM meta-categories, and one corresponding to the points in every ward in the 40 categories. We then produce two corresponding linear regression models using the log-transformed total traffic flowing into a ward as the dependent variable.

The linear regression models built with the traffic volume data show the same qualitative trend as the ones built with traffic disruption data. The first model, with the 6 meta-categories, results in an adjusted $R^2$ of $0.26$. Its coefficients indicate that OSM points tagged as \cat{commercial} are associated with heavier incoming traffic, while points tagged as \cat{recreational} are negatively associated with it. Coefficients are presented in Table S1.

The finer-grained model, featuring 40 OSM categories, naturally shows a more nuanced scenario. Not only does it provide a better fit to the data, with an adjusted $R^2$ of $0.45$, but it also provides more detail into the meta-categories used in the simpler linear models. Categories such as \cat{telephone} and \cat{university} show strong associations with higher levels of incoming traffic, whereas categories such as \cat{forest}, \cat{meadow} and \cat{allotments} show weaker associations.

\begin{figure*}[tb]
\centering
\includegraphics[width=0.9\linewidth]{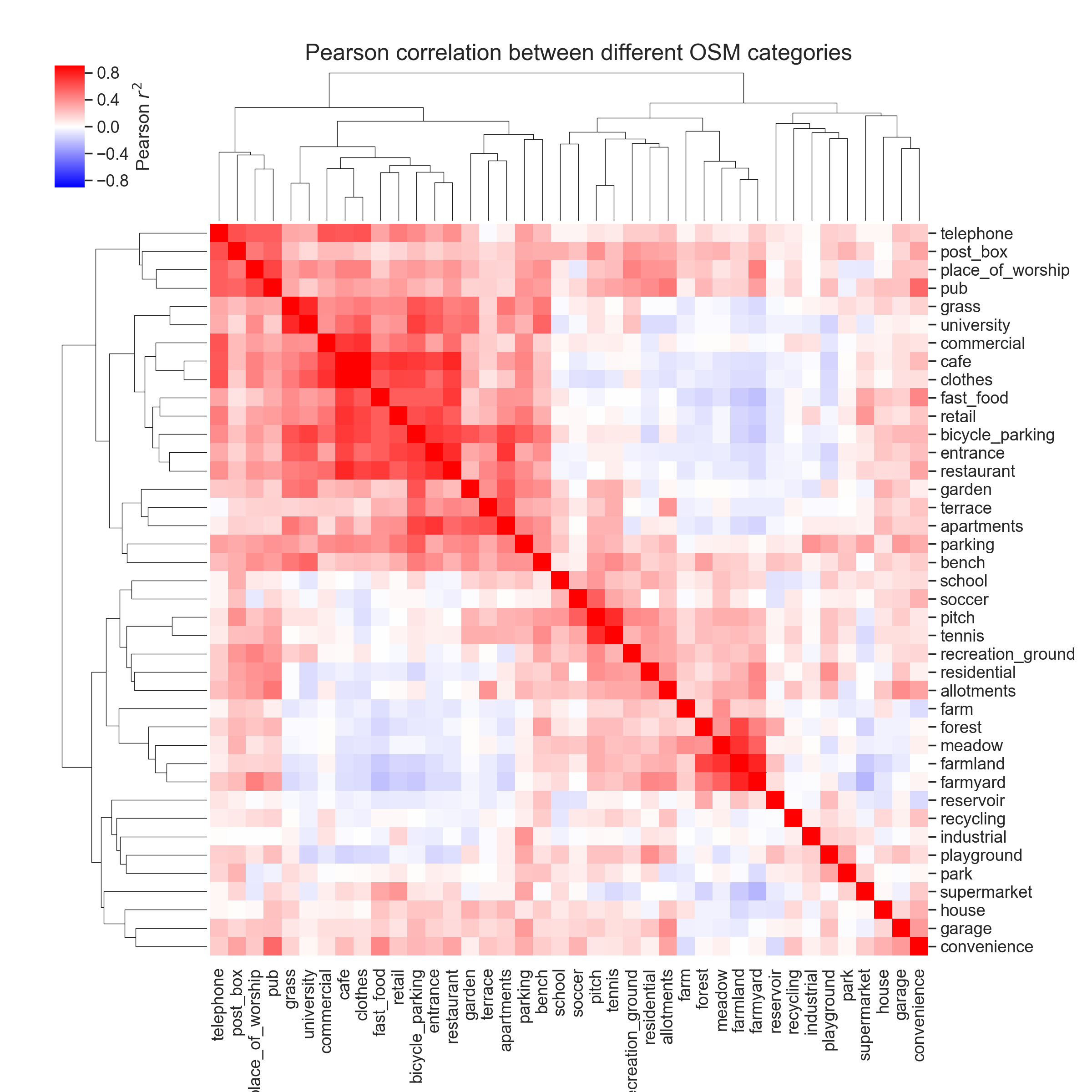}
\caption{\label{fig:clustermap}
Clustermap showing the Pearson correlation of the distribution of different OSM categories over all Oxfordshire wards. The heatmap shows the correlation between the number of points of interest tagged as every OSM category in this study. The trees show how OSM categories cluster according to their correlation. For example, OSM categories such as \cat{farm}, \cat{farmland}, \cat{farmyard} form a cluster, indicating that they often appear in the same wards, while not being as correlated to categories such as \cat{cafe} and \cat{fast food}.
}
\end{figure*}

Not surprisingly, some OSM categories are also highly correlated, in the sense that they often appear in the same wards. Figure~\ref{fig:clustermap} shows these correlations in detail. It shows a heatmap displaying the Pearson correlation between the distribution of OSM categories over wards, giving higher values to pairs of OSM categories that often appear in the same wards (e.g, \cat{forest} and \cat{meadow}), and lower values to pairs of wards that rarely co-occur (e.g., \cat{farmyard} and \cat{fast food}). The figure also shows the result of performing hierarchical clustering on the OSM categories according to their correlation. There is a cluster formed by \cat{farm}, \cat{farmland}, \cat{farmyard}, \cat{forest}, \cat{meadow}, \cat{graveyard} and \cat{reservoir}, which separates these rural categories from more urban categories as \cat{university} or \cat{retail}. 

For both the incoming traffic volume per ward and the number of traffic disruptions, the jump from 6 meta-categories to 40 OSM categories implied a change from a linear model with a poor fit to a model with a better fit, indicated by the changes in their adjusted $R^2$. It is natural to then ask if all 40 OSM categories are necessary for the new model to work, or if an equally good fit could be obtained by selecting a different number of meta-categories, or a subset of those 40 OSM categories, excluding correlated categories. This is discussed in the next subsection.

\subsection*{Feature selection}

We address the explanatory power of each variable in these linear models using feature ranking with recursive feature elimination, aided by cross-validated selection of the best number of features, as implemented in the \textit{scikit-learn} Python library~\cite{scikit-learn}. 
For both dependent variables, i.e., the incoming traffic volume per ward and the volume of traffic disruptions on a point in the road network, we perform 1000 rounds of k-fold cross-validation with $k=10$, scoring models for their $R^2$. For every cross-validation round, the 6 or 40 independent variables are then ranked according to their importance, which in this case is the magnitude of their corresponding coefficients in the linear models. Selected features are assigned rank $1$, with the next-best variable being assigned rank $2$, and so on until the last variable.

As multiple cross-validation rounds might result in different rankings of their predictor variables, we combine all rankings by calculating the stability of every variable, as well as its mean rank.
Stability selection~\cite{meinshausen2010stability} is a method which provides a useful balance between feature selection and data interpretation, by evaluating how often a given feature is included among the most important (i.e., rank $1$) for a model. Strong or important features should achieve scores close to $1$, indicating that most of the 1000 cross-validation rounds ranked them as one of the best features for prediction. Any weaker but still relevant features should still have non-zero scores, as they ought to be selected as best features at least occasionally. Finally, irrelevant features should return near-zero scores, indicating that they are very unlikely to feature among the selected variables.

For the volume of traffic disruptions, both the mean rank and the stability analysis reveal the same pattern, as shown in Tables~\ref{tbl:stability_and_ranking_LU} and~\ref{tbl:stability_and_ranking_OSM}.
The meta-category \cat{residential} features at the top, with both mean rank and stability equal to $1$, indicating a variable that featured as important in all of the 1000 cross-validation rounds. It is then followed by the meta-category of \cat{recreational}, which still features as important, with all other meta-categories featuring with a lower rank, and a stability less than $0.6$.
The corresponding granular OSM categories show the categories \cat{farmland}, \cat{residential}, \cat{parking}, \cat{forest}, and \cat{farmyard} at the top, with mean rank and stability of $1.000$, indicating that they were considered important variables in all $1000$ cross-validation rounds. These categories are followed by \cat{farm}, \cat{meadow}, and \cat{industrial}, with stability of $0.999$ and respective mean ranks of $1.001$, $1.002$ and $1.003$.

Tables~\ref{tbl:stability_and_ranking_LU} and~\ref{tbl:stability_and_ranking_OSM} also show the mean rank and stability results for the total incoming traffic volume. Reported results are for trips on weekday mornings, but qualitatively similar results are obtained when using the full collection of trips in the dataset as shown in Table S2.
The meta-category \cat{commercial} features at the top, with both mean rank and stability equal to $1$, indicating a variable that featured as important in all of the 1000 cross-validation rounds. It is then followed by the meta-category of \cat{recreational}, which still features as important, with all other meta-categories featuring with a lower rank, and a stability less than or equal to $10\%$.
The corresponding granular OSM categories show \cat{fast-food} at the top, with a mean rank and stability of $1$. The categories \cat{post box} and \cat{cafe} feature next. OSM categories such as \cat{farm} and \cat{farmyard} feature with lower mean ranks, and stability under $0.7$. 
One must bear in mind that the OSM categories \cat{residential} and \cat{commercial} are not equivalent to the meta-categories \cat{residential} and \cat{commercial}. This point is discussed in more detail in the next section.

\begin{table}[tb]
    \centering
    \subfloat[Meta-categories only, traffic disruptions]{
        \begin{tabular}{lrr}\toprule
            &               ranking	&	stability	\\	\midrule
                    residential   	&	1.000	&	1.000	\\	
                    recreational   	&	1.311	&	0.689	\\	
                    commercial   	&	1.758	&	0.553	\\	
                    industrial   	&	2.216	&	0.542	\\	
                    green space   	&	2.794	&	0.422	\\	
                    institutional  	&	3.379	&	0.415	\\	\bottomrule
        \end{tabular}
    }\hspace{5ex}
    \subfloat[Meta-categories only, traffic volume]{
        \begin{tabular}{lrr}\toprule
            &               ranking &	stability	\\	\midrule
            commercial   	&	1.000	&	1.000	\\	
            recreational   	&	1.734	&	0.266	\\	
            institutional  	&	2.676	&	0.058	\\	
            residential   	&	3.636	&	0.040	\\	
            green space   	&	4.606	&	0.030	\\	
            industrial   	&	5.602	&	0.004	\\	\bottomrule
        \end{tabular}
    }
    \caption{Average ranking and stability of different meta-categories in predicting the number of traffic disruptions and the incoming volume for every Oxfordshire ward.}
    \label{tbl:stability_and_ranking_LU}
\end{table}

\begin{table}[tb]
    \centering
    \subfloat[Granular model, traffic disruptions]{
        \begin{tabular}{lrr}\toprule
                             &          ranking	   	&	stability	\\	\midrule
                            farmland	&	1.000	&	1.000	\\	
                            residential	&	1.000	&	1.000	\\	
                            parking	    &	1.000	&	1.000	\\	
                            forest	    &	1.000	&	1.000	\\	
                            farmyard	&	1.000	&	1.000	\\	
                            farm	    &	1.001	&	0.999	\\	
                            meadow	    &	1.002	&	0.999	\\	
                            industrial	&	1.003	&	0.999	\\	
                            reservoir	&	1.010	&	0.993	\\	
                            soccer  	&	1.020	&	0.990	\\	   \bottomrule
        \end{tabular}
    }\hspace{5ex}
        \subfloat[Granular model, traffic volume]{
        \begin{tabular}{lrr}\toprule
            &           ranking			&	stability	\\	\midrule
                            fast-food	&	1.000	&	1.000	\\	
                            post box	&	1.028	&	0.972	\\	
                            cafe    	&	1.080	&	0.948	\\	
                            bench   	&	1.211	&	0.869	\\	
                            soccer	    &	1.409	&	0.802	\\	
                            commercial	&	1.648	&	0.761	\\	
                            telephone	&	1.916	&	0.732	\\	
                            parking	    &  	2.200	&	0.716	\\	
                            convenience	&	2.508	&	0.692	\\	
                            farm    	&	2.855	&	0.653	\\	\bottomrule
        \end{tabular}
    }
    \caption{Average ranking and stability of different OSM categories in predicting the number of traffic disruptions and the incoming volume for every Oxfordshire ward. Only the top 10 variables according to ranking are shown.}
    \label{tbl:stability_and_ranking_OSM}
\end{table}

\section*{Discussion}
The analysis presented in this paper shows how fine-grained land use categories can be used to estimate traffic volume and traffic disruption patterns. In particular, we have shown that the fine-grained features available on OpenStreetMap can greatly increase the explanatory power of linear models. We have also shown the importance of different land use categories by using recursive feature elimination, and have used cross-validation to examine the predictive power of different models.

One useful application of these data and methods is to offer estimated answers to questions such as ``what impact will placing another cafe at a given point have on traffic jams at that location?''. For example, according to our fine-grained traffic models, the impact of a new school on the number of traffic disruptions in its area should be comparable to the impact of a new retail store or fast food restaurant. The linear model coefficients associated with the presence of these amenities are all approximately $c_i = 0.05$, meaning that that an increase by $1$ in these variables (number of schools, retail stores, and restaurants) implies an increase of $5\%$ in the log-transformed number of traffic disruptions, i.e., an increase in $12\%$ in the monthly number of traffic disruptions at the location.
These same categories---\cat{school}, \cat{retail}, and \cat{fast food}---also have a positive correlation with the monthly volume of traffic going into a ward, even if with different coefficients. Respectively, the three categories have coefficients of $0.0010$, $0.0021$, and $0.0028$, implying respective increases in $0.2\%$, $0.5\%$, and $0.7\%$ in the total (non-log transformed) traffic flowing into areas.

It is important to remember the limitations of OpenStreetMap land use categories. For example, the OSM categories \cat{residential} and \cat{commercial} are not equivalent to the meta-categories \cat{residential} and \cat{commercial}, and the OSM dataset includes tags such as \cat{farmland} and \cat{farmyard} along with \cat{farm}, which was deprecated and substituted by the two other farm categories in 2017 \cite{deprecated-farm}.
Categories and meta-categories might differ in the quality of the annotation, and in how informative they are to the traffic predictions. The cross-validation and recursive feature elimination performed here are first steps in tackling this issue. The rank and stability analysis provide additional evidence that higher numbers of traffic disruptions are observed in residential and rural areas, indicated by meta-categories such as \cat{residential} and OSM categories such as \cat{farmland}, \cat{forest} and \cat{farmyard}. This result matches the distribution of OSM categories over all wards, as indicated in Figure~\ref{fig:clustermap}, which shows that OSM tags such as \cat{house}, \cat{farmland}, \cat{residential}, and \cat{farmyard} are often seen in the same wards, while rarely co-occurring with OSM categories such as \cat{commercial} or \cat{cafe}. The latter two OSM categories do not feature as important predictors for the number of traffic disruptions, but they do feature as important predictors for traffic volume, where they show the highest rank and stability, which is also observed for the meta-category \cat{commercial}.

Our study also suggests promising avenues for future research. One of these would be to take advantage of the constantly evolving nature of OpenStreetMap to track the emergence of new physical features, and relate these to changes in traffic conditions, thus extending the correlations we have highlighted in this paper into a causal setting. Another would be to combine these with other sources of observational data, such as licensing applications, planning permission, and building regulations, to see if these can build on the baseline model we have constructed. Finally, it would be worthwhile extending our study to other countries and contexts, to see if the value of OSM’s granular point of interest data is generalizable. As our ability to understand and explain traffic patterns improves so will the ability of policymakers to effectively design urban transport systems that serve the needs of their citizens.

\section*{Materials and Methods}

\subsection*{OpenStreetMap data}
Our geographical focus is the English county of Oxfordshire, a geographical area of just over $2,605$ km$^2$ and which contains around $680,000$ inhabitants. 
For our OpenStreetMap (OSM) data, we downloaded \cat{points of interest} from the OSM database which provide indications of the way land is used. Points of interest were downloaded in November 2017. One of the authors then assigned each point of interest to six meta-categories of land use: \cat{residential}, \cat{industrial}, \cat{commercial}, \cat{recreational}, \cat{institutional} and \cat{green space}. These categories are standard across the transport and land-use literature (see, for example, the typologies present in \cite{wegener2004land,srinivasan2013modeling,liu2012urban}). We also preserved the more granular categorization given to the points by OSM itself. For example, our meta-category of \cat{commercial} contains categories such as \cat{restaurant}, \cat{pub} and \cat{cafe}. We chose to ignore OSM categories and meta-categories with less than a hundred points of interest in Oxfordshire, as well as categories indicating the location of the transport network itself, as these are obviously coterminous with our traffic disruption data.

\subsection*{Traffic volume and traffic disruptions data}
We obtained the traffic disruption data from traffic disruption reports shared with us by the Oxfordshire County Council, which are sourced from a major traffic analytics company. These reports correspond to over $1.4$ million traffic incidents from just over 6,500 points on the Oxfordshire traffic network (each point being approximately a 10m$\times$10m square).
The number of traffic disruptions counts at each point ranged from 1 to 64,313, and with an average of 219 traffic disruption counts per point.
It is important to note that many traffic disruptions such as the ones studied in this paper do not result in casualties or police reports, meaning that data on car accidents only reflects a fraction of the incident estimates presented here. 

For the traffic volume data, we used anonymised and aggregated GPS mobile phone data provided by a major smartphone operating system.
Similar data sets have been validated and successfully used in urban mobility studies in San Francisco~\cite{sana2017using} and Amsterdam~\cite{knoop2018empirical}. The data set contains estimated trip volumes for origin-destination pairs of wards in Oxfordshire between January and February 2017 in hourly increments. We took a subset of the data, only using trips inferred by the company to be made by vehicle (and not walking or cycling), and trips on weekdays made between 7am and 12pm (noon), which we aggregated into a total traffic going into every Oxfordshire ward over the two-month period. Using the whole day and/or including weekend trips yielded qualitatively similar results.
Finally, we obtained shapefiles for the border of all Oxfordshire wards from the Digimap mapping data service \cite{digimap}. Datasets were manipulated using dataframes from the Python Pandas library \cite{mckinney2010data}.


\begin{thebibliography}{10}

\bibitem{transportbritain}
{Department for Transport}, {Transport Statistics Great Britain 2016},
  https://bit.ly/2tsCsvq.

\bibitem{vlahogianni2014short}
E.~I. Vlahogianni, M.~G. Karlaftis, J.~C. Golias, Short-term traffic
  forecasting: Where we are and where we’re going.
\newblock {\it Transportation Research Part C: Emerging Technologies\/} {\bf
  43}, 3--19 (2014).

\bibitem{mcneill2017estimating}
G.~McNeill, J.~Bright, S.~A. Hale, Estimating local commuting patterns from
  geolocated twitter data.
\newblock {\it EPJ Data Science\/} {\bf 6}, 24 (2017).

\bibitem{wegener2004land}
M.~Wegener, F.~F{\"u}rst, Land-use transport interaction: State of the art,
  http://dx.doi.org/10.2139/ssrn.1434678 (2004).

\bibitem{Lenormand2015}
M.~Lenormand, M.~Picornell, O.~G. Cant{\'{u}}-Ros, T.~Louail, R.~Herranz,
  M.~Barthelemy, E.~Fr{\'{\i}}as-Mart{\'{\i}}nez, M.~S. Miguel, J.~J. Ramasco,
  Comparing and modelling land use organization in cities.
\newblock {\it Royal Society Open Science\/} {\bf 2}, 150449 (2015).

\bibitem{Louail2015}
T.~Louail, M.~Lenormand, M.~Picornell, O.~G. Cant{\'{u}}, R.~Herranz,
  E.~Frias-Martinez, J.~J. Ramasco, M.~Barthelemy, Uncovering the spatial
  structure of mobility networks.
\newblock {\it Nature Communications\/} {\bf 6} (2015).

\bibitem{liu2012urban}
Y.~Liu, F.~Wang, Y.~Xiao, S.~Gao, Urban land uses and traffic ‘source-sink
  areas’: Evidence from gps-enabled taxi data in shanghai.
\newblock {\it Landscape and Urban Planning\/} {\bf 106}, 73--87 (2012).

\bibitem{haklay2010good}
M.~Haklay, How good is volunteered geographical information? a comparative
  study of openstreetmap and ordnance survey datasets.
\newblock {\it Environment and planning B: Planning and design\/} {\bf 37},
  682--703 (2010).

\bibitem{girres2010quality}
J.-F. Girres, G.~Touya, Quality assessment of the french openstreetmap dataset.
\newblock {\it Transactions in GIS\/} {\bf 14}, 435--459 (2010).

\bibitem{zielstra2010comparative}
D.~Zielstra, A.~Zipf, {\it 13th AGILE international conference on geographic
  information science\/} (2010), vol. 2010.

\bibitem{helbich2012comparative}
M.~Helbich, C.~Amelunxen, P.~Neis, A.~Zipf, Comparative spatial analysis of
  positional accuracy of openstreetmap and proprietary geodata.
\newblock {\it Proceedings of GI\_Forum\/} pp. 24--33 (2012).

\bibitem{mashhadi2015impact}
A.~Mashhadi, G.~Quattrone, L.~Capra, {\it OpenStreetMap in GIScience\/}
  (Springer, 2015), pp. 125--141.

\bibitem{arsanjani2015quality}
J.~J. Arsanjani, P.~Mooney, A.~Zipf, A.~Schauss, {\it OpenStreetMap in
  GIScience\/} (Springer, 2015), pp. 37--58.

\bibitem{senaratne2017review}
H.~Senaratne, A.~Mobasheri, A.~L. Ali, C.~Capineri, M.~Haklay, A review of
  volunteered geographic information quality assessment methods.
\newblock {\it International Journal of Geographical Information Science\/}
  {\bf 31}, 139--167 (2017).

\bibitem{bright2018geodemographic}
J.~Bright, S.~De~Sabbata, S.~Lee, Geodemographic biases in crowdsourced
  knowledge websites: Do neighbours fill in the blanks?
\newblock {\it GeoJournal\/} {\bf 83}, 427--440 (2018).

\bibitem{bright2018openstreetmap}
J.~Bright, S.~De~Sabbata, S.~Lee, B.~Ganesh, D.~K. Humphreys, Openstreetmap
  data for alcohol research: Reliability assessment and quality indicators.
\newblock {\it Health \& place\/} {\bf 50}, 130--136 (2018).

\bibitem{camargo2019diagnosing}
C.~Q. Camargo, J.~Bright, S.~A. Hale, Diagnosing the performance of human
  mobility models at small spatial scales using volunteered geographic
  information.
\newblock {\it arXiv preprint arXiv:1905.07964\/}  (2019).

\bibitem{choi2012predicting}
H.~Choi, H.~Varian, Predicting the present with google trends.
\newblock {\it Economic Record\/} {\bf 88}, 2--9 (2012).

\bibitem{wu2015future}
L.~Wu, E.~Brynjolfsson, {\it Economic analysis of the digital economy\/}
  (University of Chicago Press, 2015), pp. 89--118.

\bibitem{lin2019forecasting}
A.~Y. Lin, J.~Cranshaw, S.~Counts, {\it Proceedings of the 2019 World Wide Web
  Conference (WWW’19), May\/} (2019), pp. 13--17.

\bibitem{scikit-learn}
F.~Pedregosa, G.~Varoquaux, A.~Gramfort, V.~Michel, B.~Thirion, O.~Grisel,
  M.~Blondel, P.~Prettenhofer, R.~Weiss, V.~Dubourg, J.~Vanderplas, A.~Passos,
  D.~Cournapeau, M.~Brucher, M.~Perrot, E.~Duchesnay, Scikit-learn: Machine
  learning in {P}ython.
\newblock {\it Journal of Machine Learning Research\/} {\bf 12}, 2825--2830
  (2011).

\bibitem{meinshausen2010stability}
N.~Meinshausen, P.~B{\"u}hlmann, Stability selection.
\newblock {\it Journal of the Royal Statistical Society: Series B (Statistical
  Methodology)\/} {\bf 72}, 417--473 (2010).

\bibitem{deprecated-farm}
{OpenStreetMap contributors}, Openstreetmap mapnik and cartocss update,
  {https://github.com/gravitystorm/openstreetmap-carto/blob/master/CHANGELOG.md}
  (2017).

\bibitem{srinivasan2013modeling}
S.~Srinivasan, R.~Provost, R.~Steiner, Modeling the land-use correlates of
  vehicle-trip lengths for assessing the transportation impacts of land
  developments.
\newblock {\it Journal of Transport and Land Use\/}  (2013).

\bibitem{sana2017using}
B.~Sana, J.~Castiglione, D.~Cooper, D.~Tischler, {Using Google’s Aggregated
  and Anonymized Trip Data to Support Freeway Corridor Management Planning in
  San Francisco, California}.
\newblock {\it Transportation Research Record: Journal of the Transportation
  Research Board\/} {\bf 2643}, 65--73 (2017).

\bibitem{knoop2018empirical}
V.~L. Knoop, P.~B.~C. van Erp, L.~Leclercq, S.~P. Hoogendoorn, {\it 2018 21st
  International Conference on Intelligent Transportation Systems (ITSC)\/}
  (2018), pp. 3832--3839.

\bibitem{digimap}
{EDINA Digimap Ordnance Survey Service}, { OS MasterMap Topography Layer [Shape
  geospatial data], Scale 1, Tile: Oxfordshire, Ordnance Survey, Using: EDINA
  Digimap Ordnance Survey Service}, \url{ https://digimap.edina.ac.uk/ }
  (Downloaded in June 2018).

\bibitem{mckinney2010data}
W.~McKinney, {\it et~al.\/}, {\it Proceedings of the 9th Python in Science
  Conference\/} (Austin, TX, 2010), vol. 445, pp. 51--56.

\end{thebibliography}
\bibliographystyle{ScienceAdvances}

\newpage

\noindent \textbf{Acknowledgements:} The authors thank Llewelyn Morgan for facilitating access to data and supporting the project. A previous version of this paper was presented at CARMA 2018: 2nd International Conference on Advanced Research Methods and Analytics. The authors would like to thank those involved in the conference for their feedback. \\
\noindent \textbf{Funding:} This project was supported by funding from InnovateUK (grant  52277-393176), NERC (grant NE/N00728X/1), the Lloyd's Register Foundation and The Alan Turing Institute (EPSRC grant EP/N510129/1).

\noindent \textbf{Author Contributions}
SAH and JB secured the funding and coordinated the project. All authors conceived and designed the study and collected the data. CQC and JB carried out the analysis. JB wrote the first draft and CQC and SAH edited it. All authors gave final approval for publication.\\
\noindent \textbf{Competing Interests} The authors declare that they have no competing financial interests.\\
\noindent \textbf{Data and materials availability:} Additional data and materials will be available online upon publication.

\clearpage

\clearpage
\section*{Supplementary Materials}

\begin{table}[ht!]
    \centering
    \subfloat[Meta-categories only]{
        \begin{tabular}{lr}\toprule
            Variable & Estimate \\ \midrule
            Commercial & 0.0227*** \\
            Recreational & -0.0303** \\
            Institutional & 0.0087\\
            Green space & 0.0026 \\
            Industrial & -0.0005 \\
            Residential & -0.0084 \\ \midrule
            Observations & 112 \\
            Adjusted $R^2$ & 0.26 \\ \bottomrule
        \end{tabular}
    }\hspace{5ex}
    \subfloat[Granular model]{
        \begin{tabular}{lr}\toprule
            Variable & Estimate \\ \midrule
            Parking & 0.040*** \\
            House & 0.005* \\
            \ldots \\
            Pitch & -0.007\\
            Farmyard & -0.008 \\
            Pub & -0.013 \\ \midrule
            Observations & 112 \\
            Adjusted $R^2$ & 0.45 \\ \bottomrule
        \end{tabular}
    }
    \caption*{\textbf{Table S1: Coefficients for the linear regression model for the incoming traffic for every ward.} The traditional land-use categories shown in (a) that produce an adjusted $R^2=0.26$, while the granular classifications used in (b) increase the adjusted $R^2$ to $0.45$. Only a small subset of the 40 predictor variables are shown for (b).  Respectively, *, ** and *** indicate $p<0.05$, $p<0.01$ and $p<0.001$.}
    \label{tbl:SI_linreg_traffic}
\end{table}

\begin{table}[b!]
    \centering
    \subfloat[Meta-categories, total traffic volume]{
        \begin{tabular}{lrr}\toprule
            &               ranking	&	stability	\\	\midrule
            commercial	    &	1.000 	&	1.000	\\
            recreational	&	1.028	&	0.972	\\
            institutional	&	1.113	&	0.915	\\
            green space     &	1.506	&	0.607	\\
            residential     &	2.048	&	0.458	\\
            industrial      &	2.839	&	0.209	\\
	\bottomrule
        \end{tabular}
    }\hspace{5ex}
    \subfloat[Meta-categories, total traffic volume]{
        \begin{tabular}{lrr}\toprule
            &               ranking &	stability	\\	\midrule
            clothes	    &	1.000    &	1.000	\\
            bench	    &	1.000  	&	1.000   \\
            supermarket	&	1.022	&	0.978	\\
            post box	&	1.146	&	0.876	\\
            playground	&	1.317	&	0.829	\\
            fast food	&	1.525	&	0.792	\\
	\bottomrule
        \end{tabular}
    }
    \caption*{\textbf{Table S2: Average ranking and stability of different meta-categories in predicting the number of traffic disruptions and the incoming volume for every Oxfordshire ward, for trips at any time of the day.} Only the top 6 OSM categories are shown in (b). The ranking and stability results are similar to the ones obtained when only selecting trips on weekday mornings.}
    \label{tbl:stability_and_ranking_whole_day}
\end{table}

\end{document}